\documentclass[aps,prl,reprint,showpacs]{revtex4-1}

\usepackage{mathtools, amssymb, mathrsfs}
\usepackage{url}

\DeclareMathOperator{\Tr}{Tr}

\begin{document}

\title{Zermelo Navigation 
and a Speed Limit to Quantum Information Processing}
\author{Benjamin Russell}
\author{Susan Stepney}
\affiliation{Department of Computer Science, University of York, YO10 5DD, UK}
\date{\today}

\begin{abstract}
We use a specific geometric method to determine speed limits to the implementation 
of quantum gates in controlled quantum systems that have a specific class of constrained control functions.
We achieve this by applying a recent theorem of Shen, 
which provides a connection between time optimal navigation on Riemannian manifolds 
and the geodesics of a certain Finsler metric of Randers type.
We use the lengths of these geodesics to derive the optimal implementation times 
(under the assumption of constant control fields) for an arbitrary quantum operation 
(on a finite dimensional Hilbert space), 
and explicitly calculate the result for the case of a controlled single spin system in a magnetic field,
and a swap gate in a Heisenberg spin chain.
\end{abstract}


\maketitle

\section{Introduction}

There is much interest in establishing methods for determining physical speed limits for the implementation of 
quantum information processing (QIP) tasks,
both for practical engineering considerations, and for determining fundamental limits to computation.
Many of these approaches employ geometrical techniques.
Here we apply geometric methods to a specific problem, which we show to be related to the Zermelo navigation problem.
We use this method to determine a quantum speed limit (QSL) for quantum gates, in a system with pure state and a finite dimensional state space under the influence of a constrained control Hamiltonian.

\subsection{Recent work on quantum speed limits}
Recent work on the QSL falls into several categories, including:
(i) bounds on orthogonality times, (ii) time optimal quantum gates, and (iii) fundamental questions about computation.
The orthogonality time (also called passage time \cite{Brody}) is the optimal time for a system to evolve from one state to an orthogonal state.

Work on bounds on orthogonality times include \cite{Brody, MLB, Zych, UBT, GSL, CGSL, Lutz, Escher, Gab, Bend, Fring}.
Specifically \cite{Brody, GSL, CGSL} include a role for differential geometry in analysing this aspect of the QSL.
\cite{Lutz} analyses the case of an open driven system and obtains a bound also comparable to the Margolus Levitin bound for non unitary dynamics, a specific model, the damped Jayes-Cummings model is analysed.
\cite{Escher} produces an interesting result generalising the Margolus-Levitin bound to systems to systems with non-unitary dynamics.
\cite{Gab} illustrates an application of the Pontryagin minimum principle to the optimal control of $SU(2)$ operators; closed form solutions are obtained as are interesting diagrammatic representations of the optimal trajectories.
\cite{Bend} illustrates the absence of a speed limit for quantum systems described by non-Hermitian, PT-symmetric Hamiltonians in a situation where Hermitian quantum mechanics is subject to a finite speed limit.
\cite{Fring} discusses the Margolus-Levitin bound in non-Hermitian quantum systems.
Numerical methods in quantum optimal control are considered in \cite{TOQC} and the Margolus-Levitin bound is shown to be achievable using the Krotov method for deriving control schemes.
The well-known Time Energy uncertainty relation is also a bound on orthogonality time in closed, time independent systems;
a good review of this can be found in \cite{PB}, and a geometric derivation of the bound in \cite{GSL}.
A good discussion of a geometric derivation of the Time Energy uncertainty relation can be found in \cite{Brody}.

Work on time optimal quantum gates includes \cite{Li, AK, Sub-Rie, Lee, UCNC13-theory, Carlini}:
\cite{Li} discusses time optimal implementation of a number of two qubit gates and also discuss experimental implementations of such gates.
Work on open-dissipative systems for implementing quantum gates can be found in \cite{AK}.
Some works on this topic based on geometry include \cite{Sub-Rie, Lee}.
\cite{Sub-Rie} discusses the use of sub-Riemannian metrics on the unitary group with application to two and three qubit systems, special focus on NMR experiments is given.
\cite{Lee} analyses the use of metric structure (in the sense of metric spaces, not differential geometry) to determining the QSL for implementing quantum gates.
\cite{UCNC13-theory} connects the QSL for orthogonality times and the QSL for implementing quantum gates.
\cite{Carlini} produces a result based on a variational principle for a Lagrangian on $U(N)$, this work also shows how optimal control schemes can be obtained via differential geometry.

For work on the wider relevance to computer science, \cite{Lloyd, NFINCIR} are most notable.
\cite{Lloyd} discusses the role of the Margolus-Levitin bound in the context of the ultimate physical limits to computation.
\cite{NFINCIR} illustrates an application of Finsler geometry to quantum optimal control and the design of quantum circuits.

\subsection{Our geometric approach}

Here we apply geometric methods, specifically methods of Finsler geometry, to the problem of determining the QSL for quantum gates.
We impose the constraint $\Tr(\hat{H}_c^2) = 1/\alpha$ (where $\alpha$ is a positive constant) on the control Hamiltonian in a controlled quantum system;
this constraint is also considered in \cite{Carlini}.
Note that our approach is not restricted to this particular constraint, however: it would be possible to reformulate the analysis performed here if the set of allowed control Hamiltonians was the unit ball of any norm arising from an inner product on $\mathfrak{su}(N)$.
The bounds obtained here are bounds on physical times, not on any notion of circuit complexity.

Our method is based specifically on the known exact correspondence between navigation data for the problem of Zermelo navigation and Randers metrics \cite{Bao2004} (in contrast to other work applying Finsler geometry to QIP).

We provide a general solution, and then
evaluate the speed limit in the specific cases of a single spin system in a magnetic field, and a swap gate in a Heisenberg spin chain with time independent control fields.

\section{Zermelo Navigation and Randers Metrics}
Mathematically, the relevant form of the Zermelo navigation problem 
(\cite{Zermelo31}, as cited in \cite{Bao2004}) considered here comprises the following:
\begin{enumerate}
 \item A Riemannian manifold $(M, g)$, which is taken to be compact and connected.
 \item A vector field $W$ on $M$ such that $W$ is ``small'' according to the metric $g$, 
that is, $g_p(W_p,W_p) < 1$ for all points $p$ on $M$.  
In local coordinates, $g_{i j}(x) W^i(x) W^j(x) < 1$, for all points with local co-ordinates given by $x$.
\end{enumerate}
The navigator on the manifold $M$ is taken to move with unit speed according to the metric $g$.
$W$ is interpreted as a``wind'' that is ``pushing'' the navigator around, 
thus altering their speed (according to $g$) in a way that may depend on the location of the navigator 
(that is, the wind need not be the same everywhere).
The constraint on $W$ ensures that progress can always be made: 
the wind can never blow the navigator backwards.
The requirement to move with unit speed can be interpreted as ``full speed ahead'' at all times.
The problem of time optimal navigation then is to determine the direction in which 
to navigate at each point on the manifold, in order to reach some given point in minimal time.

Shen \cite{Shen} illuminates a deep connection between this problem of 
time optimal navigation on a Riemannian manifold and a specific class of Finsler metrics \cite{FINS}, 
namely, the Randers metrics.
A Randers metric is a Finsler metric that can be cast as the sum of 
a Riemannian metric and a linear term \cite{RMFC}.

Shen \cite{Shen,Bao2004} shows that, under the influence of a ``wind'', 
the time optimal trajectories are given by the geodesics of the following Randers metric:
\begin{align}
\label{navmet}
 \left\|X\right\| = & -\frac{g_p(X,W_p)}{ 1-g_p(W_p, W_p)} + {} \\[1mm]
& \frac{\sqrt{ {g_p(X,W_p)}^2 + \left(\vphantom{\strut}1-g_p(W_p, W_p)\right)g_p(X,X)}}{1-g_p(W_p, W_p)} \nonumber
\end{align}
where this formula defines the length of any tangent vector $X \in T_p M$.
Shen also shows that these geodesics lengths are the optimal times for making 
a journey between any two points on $M$.
For physical clarification: the unit sphere of the Riemannian metric encodes 
all the information about how quickly the navigator can move in a given direction 
(in the absence of any wind) at a point on $M$ by singling out the allowed tangent vectors to trajectories.
The metric is time independent throughout this work.

\section{Navigation On The Special Unitary Group}
We take Shen's result, and apply it to the case of QIP, to derive quantum speed limits.

In order to implement a certain QIP task in a controlled quantum system, 
we consider the dynamics of the system (more precisely, the time evolution operator $\hat{U}_t$) 
as given by the Schr\"odinger equation:
\begin{align}
 \frac{d \hat{U}_t}{dt} = -i\hat{H}_t \hat{U}_t = -i\left(\hat{H}_0 + \hat{H}_c(t)\right)\hat{U}_t
\end{align}
Here  $\hat{H}_t$ is the time-dependent Hamiltonian,
decomposed into the sum of $\hat{H}_0$, a ``drift'' time-independent Hamiltonian, 
representing the system's dynamics in the absence of external influences,
and $\hat{H}_c$, the control Hamiltonian that represents the effect of the 
(potentially) time-dependent influence of control fields on the dynamics.
For more details see \cite{QOC}.

In order to implement a desired computation in such a system, $\hat{U}_t$ 
(the time evolution operator acting on the system's states) 
must be driven from the identity $\hat{I}$ at $t=0$ to $\hat{O}$, 
the operator representing the desired time evolution (that is, the desired transformation of the state space).
As $\hat{U}_t$ contains the information about the dynamics of every state of the system, 
physically achieving a desired transformation of all states of a system 
is tantamount to achieving the $\hat{U}_t$ which represents this transformation.

In the case of a closed finite dimensional quantum system, 
the physical states can be identified with the set of rays in $\mathbb{C}^N$ 
(for some $N \in \mathbb{N}$); these form a complex projective space \cite{QMPHS}.
Furthermore, the set of all possible time evolutions (ignoring global phases; for more clarification of this, and a discussion of a common mathematical error, see the footnote in \cite{mese})
is the Lie group $SU(N)$, see \cite{LGBAI} for details.
That is, the set of all possible time evolution operators acting on $\mathbb{C}^N$ is $SU(N)$.

We now pose the question: when can the problem of finding optimal implementation times 
and trajectories on $SU(N)$ be posed as a special case of the Zermelo navigation problem solved by Shen?

Up to a constant multiple, there is only one bi-invariant Riemannian metric on $SU(N)$ \cite{Mil}, 
the left (or right) translation of the Killing form on $\mathfrak{su}(n)$, 
which is given by $\alpha \Tr(\hat{A}^{\dag}\hat{B})$ $\forall \hat{A},\hat{B} \in \mathfrak{su}(N)$ 
(for some fixed $\alpha \in \mathbb{R}^+$) \cite{LGBAI}.
There is no freedom to choose this aspect of the problem, except for the constant multiple $\alpha$, if bi-invariance is desired.
We use the bi-invariant metric here because of its familiarity, and because this constraint has been treated before \cite{Carlini}.
Other, right only, invariant metrics could be considered.
(See further work section for further discussion.)

One must also consider which vector field plays the role of the non-time-dependent ``wind''.
We set $\hat{W}_{\hat{U}} = -i\hat{H}_0 \hat{U}$ by examining the Schr\"odinger equation 
and observing that this vector field on $SU(N)$ describes the dynamics of the system 
in the absence of control fields.
This is the right translation of a vector at the identity, and is thus a right invariant vector field.
The form of this $\hat{W}$ is in fact simplifying when substituted into eqn.(\ref{navmet}).
In order to meet the small wind premise of the theorem 
(that is, $g_p(W_p, W_p) < 1$ for all points $p$), we require:
\begin{align}
\label{alpha_H0}
\alpha \Tr\left(\hat{H}_0^ 2\right) < 1
\end{align}

In order to meet the premise that $g$ is the metric with respect to which the navigator 
(when not affected by the wind) has velocity of exactly 1, 
we impose the following on the control Hamiltonian:
\begin{align}
\label{alpha_Hc}
\alpha \Tr\left(\hat{H}_c^2(t)\right) = 1 \quad\mbox{(for all $t$)}
\end{align}
Hence the constant $\alpha$ is determined by $\hat{H}_c(t)$.
It is in this sense that the metric arises from a physical constraint on the system's Hamiltonian; 
that is, an allowed set of Hamiltonians which are permitted to serve as tangent vectors 
to trajectories of the time evolution operator on $SU(N)$.
The unit ball of the metric at each point on the group is the set of allowed tangent vectors 
to curves and the tangent vector to a trajectory of $\hat{U}_t$ which solves 
the Schr\"odinger equation is given by $-i\hat{H}_t \hat{U}_t$ as per the Schr\"odinger equation.

Not all physical constraints need correspond to some Riemannian metric: 
the set of allowed Hamiltonians 
(which, when multiplied by $i$ could serve as tangent vectors to 
trajectories of the time-evolution operator) 
may simply not correspond to the unit balls of any Riemannian metric.
In such a case this method of Zermelo navigation would not be applicable.
Further work is needed before we can extend such a method to scenarios in which 
the manifold under consideration initially possesses a Finsler metric rather than a Riemannian one.
This would require a generalisation of Shen's theorem to Finsler manifolds.
Here we stay with the original formulation.

To summarise: we set up a navigation problem with the following elements:
\begin{enumerate}
 \item the special unitary group $SU(N)$ playing the role of the differentiable manifold $M$
 \item the metric arising (by right translation) 
from the Killing form as the Riemannian structure of this manifold
 \item the time evolution operator $\hat{U}_t$ playing the role of 
the navigator whose tangent vectors are unit vectors according to the Riemannian structure of $M$ 
 \item the drift Hamiltonian $\hat{H}_0$ playing the role of the wind $W$
\end{enumerate}

The tangent vector to any curve on $SU(N)$ at the point $\hat{U}$ is given by 
$i\hat{A}\hat{U}$ for some $\hat{A}$ satisfying $\hat{A}^{\dag} = \hat{A}$.
That is, the tangent vector is the right translation by $\hat{U}$ of some $i\hat{A} \in \mathfrak{su}(N)$.
Thus, in the special case of navigation on $SU(N)$ with $W$ as described, 
the relevant Randers metric can be shown (after some algebra) to be:
\begin{align}
\label{winmet}
 \left\|i \hat{A} \hat{U}\right\|_{\text{opt}} = & {}
 \frac{1}{\rho-1} \frac{\Tr(\hat{A} \hat{H}_0 )}{\Tr(\hat{H}_0^2)} \bigg(1 \pm {}  \nonumber \\[1mm]
& \sqrt{ 1 + (\rho-1) \frac{\Tr(\hat{H}_0^2) \Tr(\hat{A}^2) }{(\Tr(\hat{A} \hat{H}_0))^2}  } \bigg)
\end{align}
where $i\hat{A} \hat{U} \in T_{\hat{U}}SU(N)$, 
\begin{align}
\label{rho}
\rho := \frac{\Tr(\hat{H}_c^2(t))}{\Tr(\hat{H}_0^2)} > 1 
\end{align}
and the choice of $\pm$ is made to ensure positivity.
Eqn.(\ref{winmet}) depends on  $\hat{H}_c(t)$ only through $\rho$ 
($\rho$ is not time dependant as $\Tr(\hat{H}_c^2(t))$ is not time dependant, see eqn.(\ref{alpha_Hc})).
Eqn.(\ref{winmet}) has no dependence on $\hat{U}$; this metric is right-invariant.
This is a simple consequence of the fact that both $g$ and $W$ are right invariant in this application.

Note that $\Tr({\hat{H}_0}^2)$ has a fairly clear physical interpretation.
We denote the eigenvalues and corresponding eigenstates of $\hat{H}_0$ 
by $E_n$ and $|n\rangle$ respectively.
The physical meaning of this quantity can be extracted via the following derivation:
\begin{align}
 \Tr({\hat{H}_0}^2) = \sum_{n} {E_n}^2
\end{align}
Setting $|\psi_{\text{\text{un}}} \rangle = \frac{1}{\sqrt{N}} \sum_{n} |n \rangle$, 
the uniform superposition state, one observes the following:
\begin{align}
\langle \psi_{\text{un}} | {\hat{H}_0}^2 | \psi_{\text{un}} \rangle \nonumber =
\frac{1}{N} \sum_{n} {E_n}^2 = \frac{1}{N}  \Tr({\hat{H}_0}^2) \nonumber
\end{align}
and thus:
\begin{align}
 \Tr({\hat{H}_0}^2) = N \langle \psi_{\text{un}} | {\hat{H}_0}^2 | \psi_{\text{un}} \nonumber \rangle
\end{align}
which is a multiple of the expectation of ${\hat{H}_0}^2$ in the uniform superposition state.
Thus the requirement that $\alpha \Tr({\hat{H}_0}^2) < 1$ corresponds 
to constraining this physical quantity (and similarly for the control Hamiltonian).

\section{The Speed Limit}

The geodesic lengths of the metric in eqn.(\ref{winmet}) provide the speed limit 
to implementing a desired quantum gate in any quantum system meeting the premises above.
The minimal time to traverse a path from the identity $\hat{I}$ 
to some desired operator $\hat{O}$ is the length of the geodesics of eqn.(\ref{winmet}) 
connecting the two points on $SU(N)$.

As eqn.(\ref{winmet}) is a right-invariant (but not bi-invariant) 
Finsler metric on a compact connected Lie group, 
its geodesics (through the identity) are not necessarily the one parameter subgroups \cite{HGLG}.
By Stone's theorem \cite{Stone}, the one parameter-subgroups are exactly 
the curves of the form $\hat{U}_t = \exp(-it\hat{A})$ for some constant 
$\hat{A}$ such that $\hat{A}^\dag = \hat{A}$.
There may be some situations where such a curve is a geodesic; 
such geodesics are called homogeneous geodesics.
A necessary and sufficient condition for a vector in the Lie algebra of 
a connected Lie group with a left (or right, but not both) invariant Finsler metric 
to exponentiate to a geodesic is known \cite{HOMGEO}; 
investigating applications of this to QIP is the focus of further work.

\subsection{Time-independent control Hamiltonians}

In the derivation so far, the control Hamiltonian $\hat{H}_c(t)$ is time dependent 
(although $Tr({\hat{H}_c}^2)$ is time independent, eqn.(\ref{alpha_Hc})).
From now on, we restrict to cases where control Hamiltonian is not a function of time.
This results in the total Hamiltonian being time independent; 
thus all possible trajectories of the time evolution operator are 
one-parameter subgroups (generated by $\hat{A}$ say), 
since these solve the Schr\"odinger equation: $d \hat{U}_t/dt = -i\hat{A} \hat{U}_t$.

Suppose our desired operator $\hat{O}$ is reached at time $T$.
Setting $\hat{O} = \exp(-iT\hat{A})$, taking logs, and rearranging yields
$\hat{A} = \frac{i}{T}\log(\hat{O})$.
The tangent vector to a geodesic connecting the identity 
$\hat{I}$ to $\hat{O}$ is given by $\frac{i}{T}\log(\hat{O})$.
To evaluate the length $L[\hat{U}_{T}]$ of a curve $\hat{U}_{t}$ on $SU(N)$ 
according to the Finsler metric of eqn.\ref{winmet}, $\| i\hat{A} \hat{U}_t \|$, 
one integrates the length of the tangent vector to the curve along the length of the curve.
As curve lengths are independent of parametrisation, one can find the length of this curve by evaluating:
\begin{align}
 L[\hat{U}_{T}] 
&= \int_{t = 0}^{T} \left\| \frac{d \hat{U}_{t}}{d t} \right\|_{\text{opt}} d t  \nonumber \\
& =\int_{t = 0}^{T} \left\| -i \hat{A} \hat{U}_{t} \right\|_{\text{opt}} d t \nonumber \\
\end{align}
from which one obtains the optimal time:
\begin{align}
\label{bound}
& T_{\text{opt}} 
=  \frac{1}{\rho-1} \frac{i \Tr(\hat{H}_0 \log(\hat{O}))}{\Tr(\hat{H}_0^2)} \bigg( 1 \pm {} \nonumber \\[1mm]
&  \sqrt{ 1 + (\rho-1) \frac{\Tr(\hat{H}_0^2) \Tr((\log(\hat{O}))^2) }
{\left(\Tr(\hat{H}_0 \log(\hat{O}))\right)^2}}\;\; \bigg)
\end{align}
$\Tr(\hat{H}_0 \log(\hat{O}))$ is always purely imaginary, 
and thus the expression evaluates to a real result, despite the presence of $i$.
Again, the choice of $\pm$ is made to ensure positivity.

Note that this is an {\it equality} on the optimal time, not an inequality, 
under the assumptions of the problem.

We have 
\[ \lim_{\rho \to \infty} T_{\text{opt}} = 0 \]
Recall that $\rho =  \Tr(\hat{H}^2_c(t)) / \Tr({\hat{H}_0}^ 2)  > 1$, 
and that $\hat{H}_0$ is given (it is prescribed by the physics of the system) 
before any choice of $\hat{H}_c$ can be made.
So as $\rho \to \infty$, $\Tr(\hat{H}^2_c(t)) \to \infty$ necessarily.
Intuitively: as the radius of the set of allowed control Hamiltonians $\hat{H}^2_c(t)$ diverges, 
all Hamiltonians become allowed.
With no limitations on which control Hamiltonians are allowed, there is no speed limit.

Note that the constraint on the control Hamiltonian does not allow us 
to take the limit that the control Hamiltonian tends to zero 
without violating the assumption that the ``wind'' is small relative to control, 
so we cannot use it to find optimal times in the drift-only case.

Explicit comparison of this limit to existing known bounds is difficult, since the premises used to obtain eqn.(\ref{bound}) are not exactly those of any of other the known bounds cited above.
However, one can deduce what the relationship must be.
The length of any curve on $SU(N)$ gives the optimal time for traversal by $\hat{U}_t$ of a system subject to the aforementioned premises.
Thus we have found the optimal time (eqn.\ref{bound}) for traversing any trajectory achievable with time-independent controls.
Thus any other correct, comparable bound (i.e. for the same system with the same premise) must be equal to ours, or less tight.
The same goes for any bound obtained for any other trajectory of $\hat{U}_t$ by using eqn.(\ref{winmet}) and any other comparable bound as the length give the optimal time. 
This shows that Randers geometry can be used to produce exact speed limit results in driven systems; it is unknown to the authors whether or not such a bound (applying to arbitrary curves) can be obtained without the use of Randers geometry.

\section{Example I: a single spin in a magnetic field}

The result in eqn.(\ref{bound}) can be used to calculate bounds on orthogonality times in specific time-independent controlled quantum systems 
and thus assess their capacity for QIP, as in \cite{MLB}.
For concreteness, the case of a single spin in a magnetic field is used as an example.

Setting $\hat{H}_0 = B_x \sigma^{x} + B_y \sigma^{y}$ represents 
the effects of an external magnetic field outside the control of an experimenter.
Setting $\hat{H}_c = D_x \sigma^{x} + D_y \sigma^{y} + D_z \sigma^{z}$ 
represents the effects of another external magnetic field that an experimenter can control.

The requirement that $\alpha \Tr(\hat{H}_c^2) = 1$ (eqn.(\ref{alpha_Hc})) 
can be evaluated by applying the Clifford algebra 
(of $\mathbb{R}^3$ with the standard euclidean metric) 
property of the Pauli matrices $\sigma^k$ \cite{Clif}, that
$(D_k \sigma^k)^2 = (\vec{D} \cdot \vec{D}) \hat{I}$.
This implies: 
\begin{align}
\label{eqn:D}
\Tr(H_c^2)  & = \Tr(D_k \sigma^k)^2  \nonumber \\ 
& = \Tr( (\vec{D} \cdot \vec{D}) \hat{I} ) = 2 \vec{D} \cdot \vec{D} \nonumber \\ 
& = 2(D_x^2 + D_y^2 + D_z^2) = \frac{1}{\alpha}
\end{align}
Let $D^2 := D_x^2 + D_y^2 + D_z^2$.
Then $D^2 = 1/2\alpha$.

Similarly, the requirement that $\alpha \Tr(\hat{{H_0}^2}) < 1$  (eqn.(\ref{alpha_H0})) can be evaluated:
\begin{align}
\label{eqn:B}
\Tr(H_0^2)  & = \Tr((B_x \sigma^{x} + B_y \sigma^{y})^2)  = 2B^2 < \frac{1}{\alpha}
\end{align}
where $B^2 := B_x^2 + B_y^2$.
Equations (\ref{eqn:D}) and (\ref{eqn:B}) give $B^2 < D^2$; the control field overcomes the drift field.

We choose some particular operation $\hat{O}$ and calculate its optimal implementation time.
Setting $\hat{O} = $
$\begin{pmatrix}
0 & -1 \\
1 & 0
\end{pmatrix}$
gives a gate that sends each of the two computational basis states to an orthogonal state.
We then find the optimal implementation time thus:
\begin{enumerate}
 \item $\rho = D^2/{B}^2$
 \item $\log(\hat{O}) = \log \begin{pmatrix} 0 & -1 \\ 1 & 0 \end{pmatrix} 
    = \begin{pmatrix} 0 & -\frac{\pi}{2} \\ \frac{\pi}{2} & 0 \end{pmatrix} 
		= -i \frac{\pi}{2} \sigma^{y}$
 \item $\Tr\left(\log(\hat{O}) \hat{H}_0\right) = -\pi i B_y$
 \item $\Tr\left((\log(\hat{O}))^2\right) = -\pi^2/2$
\end{enumerate}

Combining these terms and substituting into eqn.(\ref{bound}) yields:
\begin{align}
 & T_{\text{opt}} = \frac{\pi}{2} \frac{B_y}{(D^2 - B^2)}\left(1 \pm \sqrt{1 + \frac{D^2 - B^2}{B_y^2}} \right)
\end{align}

When $B_y < 0$, the drift field is helping the desired operation $\hat{O}$; the $-$ve root is chosen.
When $B_y > 0$, the drift field is opposing the desired operation; 
the $+$ve root is chosen, resulting in a larger $T_{\text{opt}}$.

The optimal time depends on $D^2$, the strength of the control field and $B^2$, the strength of the external magnetic field.
The specific values of $D_x$, $D_y$, and $D_z$ (the orientation of the control field) 
that achieve this optimum time need to be calculated separately.

The metric of eqn.(\ref{winmet}) could have been used to calculate optimal times for traversing any curve in $SU(n)$, not just the time independent trajectories (i.e. one parameter subgroups) of eqn.(\ref{bound}).
These trajectories were chosen for simplicity and for their physical relevance as piecewise constant controls are frequently adopted in optimal control theory \cite{QOC}.
To find a speed limit for $\hat{U}_t$ to traverse some other curve on $SU(N)$, one would find the length of the curve according to eqn.(\ref{winmet}).

\section{Example II, A Swap Gate Implemented In A Heisenberg Spin Chain}

Another example of using eqn.(\ref{bound}) to extract a QSL is the speed limit to implementing a swap gate in a Heisenberg model spin chain.
Again the speed limit formula here refers to the optimal implementation time obtainable with constant control functions.

The matrix for a swap gate, re-phased to make it special unitary, is \cite{niel}:

\begin{align}
\hat{O} = e^{i \pi/4}
\begin{pmatrix}
1 & 0 & 0 & 0 \\
0 & 0 & 1 & 0 \\
0 & 1 & 0 & 0 \\
0 & 0 & 0 & 1 
\end{pmatrix}
\end{align}

This gate acts (upto a phase) by swapping two one qubit states: $\hat{O} |\psi_1 \rangle \otimes |\psi_2 \rangle = |\psi_2 \rangle \otimes |\psi_1 \rangle$.

The drift Hamiltonian for a two spin ``chain'' with (arbitrary spin coupling) is \cite{xyzchain}:
\begin{align}
\hat{H}_0 = \lambda_x \sigma^x \otimes \sigma^x + \lambda_y \sigma^y \otimes \sigma^y + \lambda_z \sigma^z \otimes \sigma^z
\end{align}

One easily computes the required quantities:

\begin{align}
 \log(\hat{O}) = \frac{\pi i}{4}
\begin{pmatrix}
1 & 0 & 0 & 0 \\
0 & -1 & 2 & 0 \\
0 & 2 & -1 & 0 \\
0 & 0 & 0 & 1
\end{pmatrix}
\end{align}
and thus:
\begin{align}
 \Tr((\log(\hat{O}))^2) = -\frac{3 \pi^2}{4}
\end{align}

\begin{align}
 \Tr(\hat{H}_0^2) = 4 \vec{\lambda} \cdot \vec{\lambda} =: 4 \lambda^2
\end{align}

\begin{align}
 \Tr(\hat{H}_c^2) = 1/\alpha
\end{align}
and thus:
\begin{align}
 \rho = \frac{1}{4 \alpha \lambda^2}
\end{align}
One can also compute:
\begin{align}
\Tr(\hat{H}_0 \log(\hat{O})) = \pi i (\lambda_x + \lambda_y + \lambda_z)
\end{align}
From these it follows that:
\begin{align}
& T_{\text{opt}} = \nonumber \\ 
& -\frac{\pi \alpha (\lambda_x +\lambda_y+ \lambda_z)}{1-4 \alpha \lambda^2}\left(1 \pm \sqrt{1+\frac{3(1-4 \alpha \lambda^2)}{4 \alpha(\lambda_x + \lambda_y + \lambda_z)^2} }\right)\
\end{align}
where the $\pm$ is chosen, as before, to ensure the positivity of the time.

These calculations provide some evidence that this method could be extended to three qubit gates, and perhaps higher, before intractable computations are incurred.
For three qubit gates in a similar spin chain, calculations would all be of similar length to those performed here, except for the calculation of the matrix logarithm.
Also, many good numerical algorithms exist for performing such calculations \cite{matlog} for when they become  intractable symbolically, allowing the method to be applied to much larger systems.

\section{Achieving the limit in large spin chain and spin lattice systems}

The method described in this paper assumes that any control Hamiltonian $\hat{H}_c(t)$ satisfying the constraint $\Tr(\hat{H}_c(t)^2) = 1/\alpha$ can be implemented.
For large quantum systems, specifically larger spin chains and lattices, this will almost never be the case.

For example, in the three spin chain case, consider the term $\sigma^{x} \otimes \hat{I} \otimes \sigma^{x}$. This term represents the spin-spin coupling of non-neighbouring spins, an interaction term that produces dynamics not equivalent to that produced by any external field.
Thus any optimal times for such a model calculated using this approach would be theoretical optimal times only.
They would provide only a bound on speeds limits achievable by physically possible control Hamiltonians.
The authors do not know of a driven, finite dimensional quantum system for which $\sigma^{x} \otimes \hat{I} \otimes \sigma^{x}$ is a plausible term in the control Hamiltonian.
(See the future work section for discussion of modifications of the method taking into account systems the are not completely controllable.)

\section{Summary}

We have obtained a closed form expression eqn.(\ref{bound}) for the optimal implementation times 
for an arbitrary quantum operation on a finite dimensional Hilbert space 
in the presence of a specific constraint on the time-independent control Hamiltonian: 
that it is constant in size (in the specific sense above), and stronger than the drift Hamiltonian.

We have done this by finding a Randers metric with a special property.
The metric of eqn.(\ref{winmet}) has the property that the length of \emph{any} curve on $SU(N)$ is the optimal traversal time (for $\hat{U}_t$) for a quantum system subject to the constraints discussed.
This is in contrast to other methods that typically compute the optimal time for the optimal trajectories, or for only the trajectories achievable with time-independent Hamiltonians.
Our method applies to all trajectories whose lengths can be computed.
Finding the geodesics of eqn.(\ref{winmet}) would find the globally (over all paths with fixed endpoints) time optimal trajectories: however, these curves may not be trajectories achievable with time-independent Hamiltonians.

\section{Further work}

The examples illustrate the method in the case of constant controls.
It appears that the calculations, at least in the case of constant controls, are tractable by hand for a variety of 2 and 3 qubit gates.
The obstacle to applying the result to much larger system will be the calculation of matrix logarithms of large matrices, especially in cases when there are interaction terms in the Hamiltonian.

We plan to solve the geodesic equation for the metric in eqn.(\ref{winmet}) in order to determine speed limits for traversing more general curves than the time independent trajectories alone.
This would produce an explicit QSL formula for any trajectory of $\hat{U}_t$ with a time dependant Hamiltonian as the length of any curve according to eqn.(\ref{winmet}) gives the optimal time for $\hat{U}_t$ to traverse it in the presence of the constraint discussed.
Finding the geodesic vectors and thus the homogeneous geodesics of eqn.(\ref{winmet}) would determine exactly when time independent controls are in fact time optimal and we hope to do this using the results in \cite{HOMGEO} and elsewhere.
 
We also intend to generalise to navigation problems on $SU(N)$ where the metric representing the constraint is a Finsler metric rather than a Riemannian one.
This work will be based on the general formalism in \cite{Shen}, particularly \S3:eqn.(12) and following results.
We will also investigate using a different Riemannian metric to start with, that is, a different physical constraint.
The right translation of any inner product on $\mathfrak{su}(N)$ would produce such a metric, so there is a rich source of example quadratic constraints that can be studied this way.
The investigation of the geodesics of general right invariant Randers metrics on $SU(N)$ can be approached by applying the Euler-Poincar\'{e} equation \cite{EP}, which should provide a first order differential equation satisfied by the optimal Hamiltonian (the one that drives the $\hat{U}_t$ along a geodesic).

Lagrange multiplier methods can be used to further constrain the control Hamiltonian so that some terms in the control, i.e. the ones that physically cannot be implemented, are set to zero along trajectories.

We are also investigating the use of Koprina metrics \cite{Yoshikawa2012}, which provide other solutions to the navigation problem under different assumptions.
These metrics correspond to the case of the drift and control Hamiltonians being equal is size and thus could facilitate an analysis of the potential of low power quantum devices.
 
\begin{acknowledgments}
Our thanks to Sam Braunstein and Tony Sudbery for helpful discussions and comments, 
and to Eli Hawkins for introducing us to Finsler geometry.
Russell is supported by an EPSRC DTA grant.
We also acknowledge the helpful comments of anonymous referees.
\end{acknowledgments}

\bibliography{p3}

\end{document}